\documentclass[man,floatsintext]{apa7}
\usepackage{graphicx}%
\usepackage[labelsep=newline,labelfont=bf]{caption}
\usepackage{csquotes}
\usepackage{booktabs}
\usepackage{tabularx}
\usepackage{multirow}
\usepackage{longtable}
\usepackage{threeparttable}
\usepackage[style=apa,sortcites=true,sorting=nyt,backend=biber]{biblatex}
\DeclareLanguageMapping{american}{american-apa}
\title{Mapping the Gender Attrition Gap in Academic Psychology}
\shorttitle{Gender Attrition Gap in Academic Psychology}

\authorsnames[1,1,1,{1,2}]{Xinyi Zhao, Anna I. Thoma, Ralph Hertwig, Dirk U. Wulff}

\authorsaffiliations{{Center for Adaptive Rationality, Max Planck Institute for Human Development, Berlin, Germany},{Center for Cognitive and Decision Science, University of Basel}}

\authornote{
   Harris Cooper served as action editor.

   \addORCIDlink{Xinyi Zhao}{0000-0002-2552-7795}
   \addORCIDlink{Anna I. Thoma}{0009-0009-8991-6202}
   \addORCIDlink{Ralph Hertwig}{0000-0002-9908-9556}
   \addORCIDlink{Dirk U. Wulff}{0000-0002-4008-8022}

   This article was funded by the German Science Foundation (Grant SPP 2317) awarded to Dirk U. Wulff.

    We thank Robin Haunschild for providing access to the Scopus data and Deb Ain for editing the manuscript. This work was supported by the German Research Foundation (DFG) through Priority Programme SPP 2317 (Grant 546419617), awarded to Dirk U. Wulff, and Grant HE 2768/11-1, awarded to Ralph Hertwig. The authors declare no conflicts of interest.

   Xinyi Zhao played a lead role in conceptualization, data curation, formal analysis, methodology, visualization, writing--original draft, and writing--review and editing. Anna I. Thoma played a supporting role in conceptualization, writing--original draft, and writing--review and editing. Ralph Hertwig played a supporting role in writing--review and editing. Dirk U. Wulff played a lead role in funding acquisition and supervision and a supporting role in conceptualization, methodology, writing--original draft, and writing--review and editing.

   Correspondence concerning this article should be addressed to Xinyi Zhao, Center for Adaptive Rationality, Max Planck Institute for Human Development, Lentzeallee 94, 14195 Berlin, Germany. Email: zhao@mpib-berlin.mpg.de}

\abstract{Women comprise the majority of students and early-career scholars in psychology, yet they are less likely to remain active in research over time. This pattern raises a central question: At what stages of academic careers do women disproportionately leave academia, and what factors drive their attrition? Using large-scale bibliometric data tracking 78,216 psychologists who began publishing between 2000 and 2014, we examine gender differences in research career attrition operationalized through publishing activity across the full trajectory from entry onward. Although women accounted for more than 60\% of new entrants, they experienced higher attrition rates than men, with the gender gap peaking approximately 5 years after first publication. Early-career performance---particularly first-authored publications---was the strongest predictor of subsequent retention, whereas last-authored publications were most closely associated with continued activity at later-career stages. Collaboration patterns and institutional context also shaped career persistence, though less strongly than publication indicators. Notably, gender differences in research attrition persisted even after accounting for these career determinants, especially during early-career stages. These findings suggest that gender inequality in psychology is driven less by recruitment than by differential retention over time. Addressing early-career vulnerability may therefore be essential to achieving equitable representation in senior academic leadership within the discipline.
}


\begin{document}
\maketitle

\section*{Public Significance Statement}
Women enter psychology in large numbers but are less likely to remain active in research over time than men. Using publication records to track research continuity over time, this study shows that early-career attrition, rather than recruitment alone, contributes to persistent gender gaps at senior levels. Supporting women in the early stages of academic careers may be critical for achieving equitable representation in the discipline.


\clearpage
Gender disparities in academia are well-documented \parencite{EC_shefigure_2021, Elsevier17Gender, Nielsen_diversity_2018}---particularly in traditionally defined science, technology, engineering, and mathematics fields, where women are underrepresented from graduate education through to senior leadership positions \parencite{Blackburn_stem_2017, EC_shefigure_2021,Kwiek2025_attrition}. Prior work has emphasized entry-level barriers as a key driver of gender inequality in academia \parencite{Etzkowitz2000_women, Blackburn_stem_2017, EC_shefigure_2021}. However, this focus may obscure a distinct dynamic in women-dominated disciplines. In fields such as psychology and the social sciences more broadly, inequality is less about access and more about retention: Women's numerical dominance at junior levels fails to translate into parity at the top \parencite{Casad2022_SocialScience, LaBerge_attrition_retention_2024}. Women have consistently outnumbered men among psychology undergraduate and graduate students for decades and now constitute three quarters of psychology students \parencite{Gruber_psychology_2020, NSF_psychology_2018}, but only about a third of full professors are women \parencite{Larivire_gender_2013, Odic_pubgap_2020}. The sharp decline from women’s relatively high initial recruitment to their minority presence in professorships reflects a leaky pipeline, but one that differs from other scientific disciplines, where male dominance is evident from the earliest stages. A key unanswered question is therefore when women disproportionately exit research careers in psychology. This study contributes to this question by tracing gender differences in academic attrition over the course of academic careers, examining the factors associated with these differences, and assessing whether patterns of attrition have changed over recent decades.

Despite persistent gender disparities at senior ranks, research on academic advancement within psychology presents a comparatively optimistic picture. Several studies report that gender gaps in tenure attainment and time-to-tenure have narrowed substantially in recent decades \parencite{Gruber_psychology_2020}. Analyses of U.S. faculty data suggest that, conditional on holding tenure-track positions, women and men in psychology increasingly exhibit comparable probabilities of achieving tenure and similar timelines for promotion \parencite{Janet_survival_social_2015, webber2018there}. These findings have been interpreted as evidence that formal evaluation and promotion processes within psychology may be becoming more gender equitable. However, parity in promotion rates conditional on remaining in academia does not necessarily translate into equal long-term representation at senior ranks. If women disproportionately exit academia before, between, or after key promotion milestones, cumulative attrition can still generate substantial vertical segregation, even in the absence of large promotion gaps. Indeed, broader bibliometric and cross-disciplinary research continues to document women’s underrepresentation among full professors, highly cited scholars, and academic leaders \parencite{Larivire_gender_2013, Elsevier17Gender, Nielsen_diversity_2018,Casad2022_SocialScience, Kwiek2025_attrition}. Moreover, much of the empirical evidence on tenure and promotion is based on institutionally bounded data, often limited to specific national context \parencite{Janet_survival_social_2015, webber2018there, Gruber_psychology_2020, Casad2022_SocialScience, LaBerge_attrition_retention_2024, Habicht2026_leaky_pipeline}, limiting its generalizability across national systems that differ substantially in promotion structures, funding regimes, and work–family policies \parencite{EC_shefigure_2021}. An international, career-long perspective that examines how gender differences in academic attrition unfold across the full career trajectory may help inform targeted interventions that support women's retention and progression in academic psychology internationally.

Academic performance, collaboration networks, and institutional affiliations are well recognized as central to career continuity and academic advancement \parencite{Ceci_causes_2011, Larivire_gender_2013, vanderWal2021_collaboration, Wapman2022_hiring, Kwiek2023_promotion}, but remain to be explored in more detail in psychology \parencite{West2013_authorship, Gruber_psychology_2020,Odic_pubgap_2020}. Key measures of academic performance, such as scientific productivity, authorship position, and citation impact, directly influence hiring and promotion decisions \parencite{Kwiek2023_promotion}. In psychology, the gender publication gap has been documented while controlling for different contexts (e.g., institutional affiliation or subfields) and remains particularly evident at the last-author position---suggesting continued challenges for female researchers at senior levels \parencite{Odic_pubgap_2020}. Large and diverse collaboration networks benefit career progression, particularly in the early stages of a researcher's career \parencite{vanderWal2021_collaboration, Paraskevopoulos2021_collaboration}. Finally, institutional prestige shapes access to resources, mentorship, and funding \parencite{Clauset_hiring_2015, Gumpertz_Retention_2017, Wapman2022_hiring, Spoon_retention_2023}. Women exhibit systematic differences across these dimensions compared with men, with important implications for career progression, including lower chances of securing tenure and promotion to leadership positions \parencite{Ceci_causes_2011, Larivire_gender_2013, Schmaling2023_gender}. Critically, the extent to which each of these factors contributes to attrition in psychology is currently unknown. Moreover, beyond these observable factors, structural barriers such as systemic bias in evaluation processes and disproportionate caregiving responsibilities may also shape women's career trajectories yet often fall outside the scope of large-scale, data-driven studies \parencite{Gumpertz_Retention_2017, Guarino_service_2017, Mason_2013_baby, Kwiek2025_attrition}. A key question motivating this study is, therefore, the extent to which gender differences in academic attrition can be explained by observable career characteristics.

To address these limitations, this study adopts a complementary approach by focusing on career continuity in research output. Because comprehensive data on academic rank and promotion are not consistently available at scale, we rely on publication-based indicators, which are widely used in the literature as proxies for research engagement and career persistence in large-scale, cross-national studies \parencite{Kwiek2025_attrition, GonzalezSalmon2025_bibliometric}. Accordingly, we conceptualize academic careers in terms of sustained research activity and define attrition as the cessation of publishing over time. Using large-scale bibliometric data, we track researchers longitudinally from their entry into publishing in psychology, enabling a global and career-long perspective on retention. This approach makes it possible to identify when gender differences in persistence emerge and to examine how they accumulate over time while clearly distinguishing publication-based career trajectories from formal academic rank.

Our research aimed to advance the understanding of gender disparities in academic psychology by examining when gender differences in academic attrition arise, how they change across career stages, and the extent to which they can be explained by differences in research performance, collaboration networks, and institutional context. By providing a longitudinal, career stage--specific perspective, the study offers a nuanced insight into the leaky pipeline in academic psychology and provides evidence that can inform targeted, career stage--specific strategies to promote gender equity in academic progression.

\section*{Method}\label{sec11}

To examine gender differences in research career continuity, this study relies on large-scale bibliometric data from Scopus, one of the most comprehensive and widely used academic databases, which provides extensive coverage of scientific publications and citation networks. Our data processing pipeline (see Supplemental Figure~S1) involved identifying researchers in psychology, reconstructing their publication-based career trajectories, and extracting key factors related to research activity and persistence. We began with 1,479,461 psychology publications indexed in Scopus, corresponding to 4,965,084 unique authors who had published at least one article in a psychology journal.

\subsection*{Data Processing}
\subsubsection*{Psychology Entrants}

We identified \textit{psychology entrants} as researchers who (a) had more than one career publication and (b) published at least 60\% of their articles in psychology---as categorized by Scopus---during their first 3 academic years. Applying these criteria yielded 91,774 psychology entrants.

Scopus uses the All Science Journal Classification to organize psychology into eight subfields: general psychology, clinical psychology, developmental and educational psychology, social psychology, applied psychology, experimental and cognitive psychology, neuropsychology and physiological psychology, and psychology (miscellaneous). Publications can fall under multiple categories. We assigned each researcher a primary subfield based on the most frequent category in their first 3 years. In cases of a tie, we prioritized the subject of their first publication. When general psychology was among the most frequent categories, we prioritized a more specialized field whenever possible.

Scopus provides reliable coverage of publication data from 1996 onward \parencite{Falagas2007_scopus,Mongeon2015_scopus}. To mitigate left truncation bias, whereby more senior researchers might be misclassified as new entrants, we restricted our sample to researchers whose first Scopus-indexed publication occurred between 2000 and 2014. Limiting the entry window to 2014 ensured at least 10 years of observable career data for each researcher. 

In addition to Scopus’s built-in author disambiguation procedures, we implemented further cleaning steps to identify duplicate individuals and potentially misassigned Scopus IDs (see the Supplemental Material for more details). These procedures included consolidating duplicate profiles, correcting likely ID conflations, and excluding ambiguous entrant records. In total, 3,574 profiles were removed through disambiguation and data-quality checks, resulting in a final data set of 88,200 disambiguated psychology researchers. Disaggregated statistics of female and male psychology entrants by subfield and cohort group are presented in Supplemental Table~S1.

\subsubsection*{Academic Trajectories}
We defined a researcher's \textit{academic age} as the number of years since their first publication, using publication activity as an observable indicator of research career participation. The year of the first publication was recorded as a researcher's \textit{cohort}, with our data set covering cohorts from 2000 to 2014. To facilitate analysis while accounting for longitudinal trends, we grouped cohorts into three broader \textit{cohort groups}: 2000--2004 (researchers who began publishing between 2000 and 2004), 2005--2009, and 2010--2014. This aggregation enables clearer comparisons of publication-based career trajectories across cohorts.

To operationalize academic attrition (i.e., exit from active research output), we applied a 5-year publication gap rule: A researcher was considered to have exited when no publications were observed in the subsequent 5 years. Given that our data set extends through July 2024, researchers whose last publication occurred in 2018 or earlier were classified as having exited in the year of their final publication. Researchers with publications after 2018 were treated as right censored, as their subsequent publication activity cannot be fully observed within the data window.

\subsubsection*{Gender Detection From First Names}
To infer the gender of researchers, we used Genderize.io, a large database that links first names to binary gender categories (female/male). After basic text cleaning, such as removing entries that contained only initials, we assigned gender only when Genderize.io returned a probability score greater than 0.7; otherwise, no gender was assigned. Of the 82,917 researchers with identifiable gender (after excluding 5,283 unclassified cases), 51,143 (61.68\%) were inferred as women and 31,774 (38.32\%) as men. Finally, 78,216 psychology entrants with identifiable subfield, affiliation, and country information were retained for further analysis (after excluding 4,701 cases with missing data), comprising 48,611 women (62.16\%) and 29,605 men (37.84\%).

\subsection*{Career-Related Factors}

To examine the academic trajectories of psychology entrants, we analyzed the temporal development of several metrics across three domains: academic performance, collaboration networks, and institutional backgrounds. 

\subsubsection*{Academic Performance}
Researchers’ academic performance at each academic age was assessed using three key indicators: (1) the citation scores of their publications, (2) the number of first-authored publications, and (3) the number of last-authored publications. Each publication's citation scores were measured by the accumulated 5-year Discipline- and Year-Normalized Citation Score (DNCS). This score is calculated by dividing a publication’s actual citation count over the 5 years following publication by the average citation count of all articles published in the same discipline and years. This normalization accounts for disciplinary and temporal variations in citation practices, thereby enabling fair comparisons of scientific impact across fields and cohorts. Additionally, to capture individual contribution in research outputs, we tracked the annual number of first-authored publications, where the author is the main contributor, and last-authored publications, which typically reflect seniority. Together, these indicators help assess a researcher’s scientific independence and hierarchical progress over time.

\subsubsection*{Collaboration Network}

The characteristics of researchers’ collaborative networks were assessed by the networks' size, gender composition, and scientific impact using three metrics: the number of unique coauthors, the proportion of female coauthors, and the average scientific impact within the collaboration network at each academic age. These metrics capture both the breadth and quality of collaborative relationships over time. For each researcher and academic year, we calculated the proportion of female collaborators. To quantify the scientific impact of collaborators, we averaged the DNCS of each coauthor for the 5 years preceding a collaboration and assigned this value as the impact of a researcher's collaborative network for the year of the collaboration. This measure reflects the average scientific influence of a researcher’s collaboration networks over time. 

\subsubsection*{Institutional Environment}
We used three dimensions to assess the institutional environment: the region of original affiliation, the number of unique affiliations, and the prestige of affiliated institutions (measured by the 5-year DNCS of psychology publications). Each entrant’s affiliation was determined annually based on the institutions listed in their publications, with the region of origin defined by the country of their primary institution in the first publication year. The psychology entrants were concentrated in the United States: 40.1\% of entrants were affiliated with U.S. institutions at their first publication, followed by 8.4\% with British institutions, 5.4\% with German institutions, and 5.3\% with Canadian institutions. Around 19\% of entrants were affiliated with countries outside of the United States, Canada, and Western Europe. This skewed distribution is consistent with prior findings on the dominance of a few Western countries in global psychology research \parencite{Arnett2008_americanPsy, Tindle2021_GeographyPsy}.  

The number of unique affiliations at each academic age was used as an indicator of academic resources, and institutional prestige was quantified by the cumulative 5-year DNCS of psychology publications indexed in Scopus. For researchers with multiple affiliations in a given year, we assigned the score of the highest ranked institution. This design captures the evolving institutional environment surrounding researchers across their careers.

\subsection*{Transparency and Openness}
Data and code are available on GitHub at \url{https://github.com/zxy919781142/Sustainable-Development-Goals-in-Psychology/tree/main}. The analyses reported in this article were not preregistered.

\section*{Results}\label{sec2}
Table~\ref{tab:summary} provides a consolidated overview of the key quantitative findings reported across this section, organized by domain of analysis. Detailed descriptions of each result, including the factors associated with attrition and the role of gender across career stages, are provided in the subsections below.

\begin{table}[!htbp]
\renewcommand{\baselinestretch}{1}\scriptsize
\renewcommand{\arraystretch}{1.2}
\caption{Summary of Key Quantitative Results}
\label{tab:summary}
\begin{threeparttable}
\begin{tabular}{@{}p{2.8cm} p{3.6cm} p{7.6cm} p{1.4cm}@{}}
\toprule
Domain/analysis & Metric & Finding & Source \\
\midrule
Entry and composition
  & Female share of entrants
  & Greater than 60\% by 2014; range = 57.8\% (applied psychology) to 77.1\% (developmental and educational psychology)
  & Figure~\ref{fig:career_cohort}A \\
  & Cohort growth 2010--2014
  & Women approx.\ 2$\times$, men approx.\ $+$40\%
  & \\
\midrule
Survival (KM)
  & Survival probability, age 5
  & 0.70 F versus 0.76 M
  & Figure~\ref{fig:career_cohort}B \\
  & Survival probability, age 10
  & 0.54 F versus 0.60 M
  & \\
  & Survival probability, age 15
  & 0.47 F versus 0.53 M
  & \\
  & Median career length
  & approx.\ 13-year F versus approx.\ 18-year M
  & \\
\midrule
Transition rates
  & Transition rate, age 5
  & 0.908--0.934 F versus 0.923--0.951 M
  & Figure~\ref{fig:career_cohort}C \\
  & Early-career gap (age 4--6)
  & Approx.\ 1.5\% points per year favoring men, all cohorts
  & Figure~\ref{fig:career_cohort}D \\
\midrule
Profile gaps
  & Citation scores
  & Largest difference approx.\ 0.4 points at age 7, favoring men
  & Figure~\ref{fig:var_stat} \\
  & No.\ of first-authored articles
  & Largest difference approx.\ 0.1 articles at age 7, favoring men
  & \\
  & No.\ of last-authored articles
  & Largest difference approx.\ 0.2 articles at age 12, favoring men
  & \\
  & Female coauthor proportion
  & Largest difference approx.\ 0.08 at age 15, favoring women
  & \\
  & Affiliation prestige
  & Largest difference approx.\ 150 points at age 5, favoring women
  & \\
\midrule
Attrition predictors (TVEM)
  & Top predictor
  & No.\ of first-authored articles (peaks age 4--5), all cohorts
  & Figure~\ref{fig:importance} \\
  & Second predictor
  & No.\ of last-authored articles (late-career)
  & \\
  & Citation scores
  & Strongest in 2010--2014 cohort
  & \\
  & Gender effect
  & Persistently increases attrition risk across all career stages and cohorts
  & \\
  & Coauthor impact
  & Retention benefit strongest in early career, diminishes over time
  & \\
\midrule
Adjusted gender gap
  & 2000--2004
  & Peak gap 5\% points at age 3 (20\% F vs.\ 15\% M); closes approx.\ age 15
  & Figure~\ref{fig:prediction} \\
  & 2005--2009
  & Approx.\ 2\% points, persistent
  & \\
  & 2010--2014
  & Approx.\ 3\% points from age 2, stable through age 9
  & \\
\bottomrule
\end{tabular}
\begin{tablenotes}
\renewcommand{\baselinestretch}{1.3}\footnotesize
\vspace{4pt}
\item \textit{Note.} Profile gap values reflect the largest observed mean difference between men and women and the academic age at which this difference peaks. Adjusted gender gap refers to predicted attrition risk holding performance, collaboration, and institutional factors constant. approx.\ = approximately; F = female; M = male; KM = Kaplan--Meier; yr = years; No.\ = number; TVEM = time-varying effects model.
\end{tablenotes}
\end{threeparttable}
\end{table}

\subsection*{Women Have Shorter Academic Careers and Higher Early-Career Attrition in Publication Records}\label{result1}

\begin{figure}[htbp!]
\centering
\caption{Gender Differences in Academic Career Trajectories}
\includegraphics[width=1\textwidth]{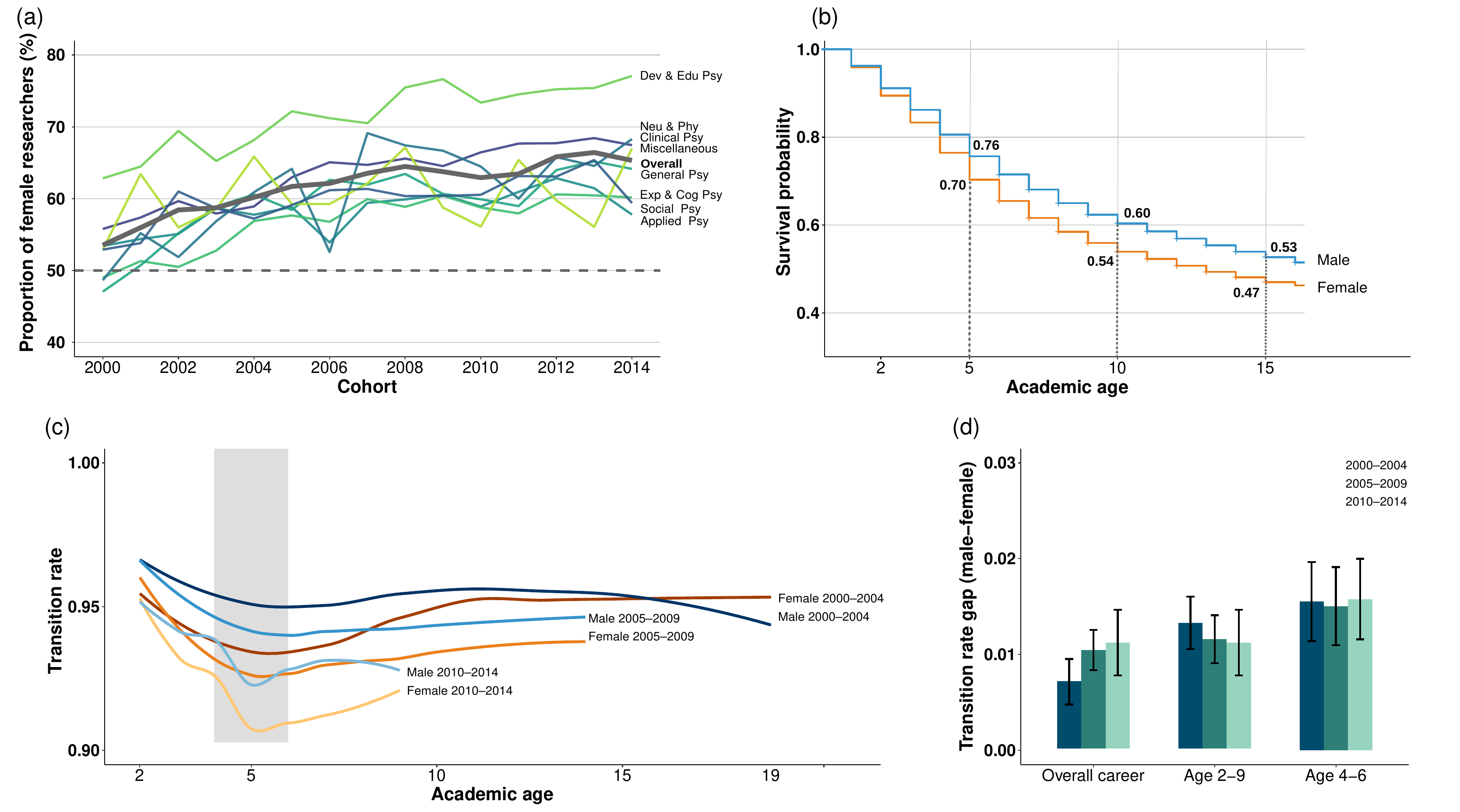}
\label{fig:career_cohort}
\vspace{6pt}
\begin{minipage}{\textwidth}
\renewcommand{\baselinestretch}{1.3}\footnotesize
\textit{Note.} (A) Proportion of women among psychology entrants, by subfield and cohort. (B) Kaplan--Meier survival probability of psychology entrants by gender. (C) Annual transition rates (i.e., the probability of continuing publishing  from the previous year) by academic age and cohort group (2000–2004, 2005–2009, 2010–2014). The gray bar indicates years in which retention dropped most sharply. Note that cohort groups differ in their career lengths as a result of the data availability. (D) Summary of gender gaps in transition rates across three time frames: the full captured career span (which varies by cohort), a shared-career window common to all cohort groups (academic age = 2–9), and the early-career period (academic age = 4–6). Positive values indicate higher transition rates for men. Dev \& Edu Psy = developmental and educational psychology; Neu \& Phy = neuroscience and physiological psychology; Clinical Psy = clinical psychology; General Psy = general psychology; Exp \& Cog Psy = experimental and cognitive psychology; Social Psy = social psychology; Applied Psy = applied psychology.
\end{minipage}
\end{figure}

We examined the proportion of female psychology entrants across subfields and cohorts, observing a substantial increase in female representation over time (see Figure~\ref{fig:career_cohort}A). In 2000, women already outnumbered men overall in psychology. Between 2010 and 2014, the number of women nearly doubled, whereas the number of men increased only about 40\%. This disproportionate growth led to an increasingly unbalanced gender composition, resulting in a widening gender gap. By 2014, women constituted over 60\% of new psychology researchers. However, there are differences between subfields, for instance, nearly 75\% of researchers in developmental and educational psychology were women, while this percentage was only 58\% in applied psychology.

Despite women entering academic careers in psychology at disproportionately higher rates, we found that women's academic careers were, on average, shorter than men's. Kaplan--Meier survival estimates showed that women were, overall, less likely to remain actively publishing  (see Figure~\ref{fig:career_cohort}B). Specifically, the estimated survival probability at academic age 5 was 0.70 for women compared with 0.76 for men, and this gap remained similar at age 10 (0.54 vs. 0.60) and age 15 (0.47 vs. 0.53). Consistent with these differences, around 51\% of women in psychology had ceased publishing by academic age 13, whereas half of men did not leave until academic age 18. 

To evaluate when gender gaps were most pronounced, we examined annual transition rates, the probability of remaining active from the previous year to each target year, by gender and cohort group (see Figure~\ref{fig:career_cohort}C; see the Supplemental Material for formal definition). For both genders, retention dropped most sharply in the early-career years, between academic ages 4 and 6. At academic age 5, fitted transition rates ranged from 0.908 to 0.934 for women, compared with 0.923 to 0.951 for men across cohort groups, showing that women's transition rates were consistently lower than men's throughout this critical period. After academic age 6, women’s transition rates gradually recovered at a faster pace than men’s, especially in the 2000–2004 cohort group, which has the longest observed career span. Women who persisted in publishing beyond the early-career stage showed higher retention rates than men in later years, surpassing them after academic age 16.

Complementing these trends, Figure~\ref{fig:career_cohort}D shows average gender gaps in transition rates across three career windows: the full observed career span (which varies by cohort due to data availability), a shared window common to all cohort groups (academic age = 2–9), and the early-career period with the lowest transition rates (academic age = 4–6). Because gender gaps tended to narrow beyond the early-career stage, the smallest average gap in transition rates across the full captured career span was found in the 2000–2004 cohort group. By contrast, during the shared-career period (age = 2–9), gender gaps decreased across successive cohorts, potentially signaling some progress toward gender equity at comparable stages of the academic trajectory. However, this progress was less evident during the critical early-career period (academic age = 4–6), where transition rates dropped sharply for both genders. We found a persistent gender gap of approximately 1.5 \% points per year in favor of men during this period across all three cohort groups. While this difference may appear small in a short time frame, even modest disparities, when applied to large researcher populations and accumulating year after year, can compound into substantial inequalities in sustained research over the course of academic careers.

\subsection*{Gender Differences in Academic Profiles: Men Lead in Publishing Performance and Women in Collaboration and Prestige}\label{result2}

\begin{figure}[htbp!]
\centering
\caption{Statistical Description of Factors Associated With Academic Performance, Collaboration, and Institutional Affiliations Across Career Stages, Disaggregated by Gender}
\includegraphics[width=1\textwidth]{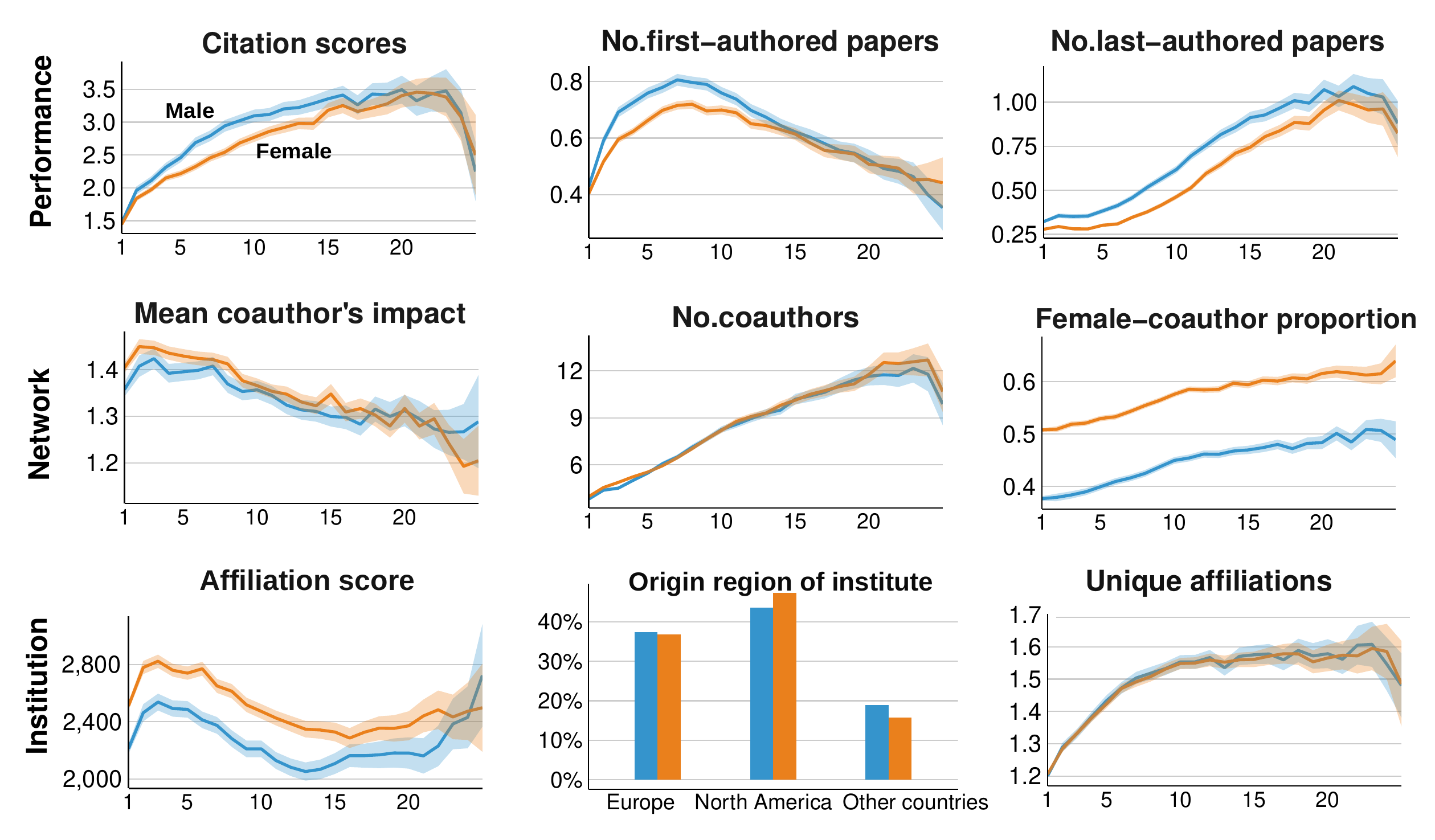}
\label{fig:var_stat}
\vspace{6pt}
\begin{minipage}{\textwidth}
\renewcommand{\baselinestretch}{1.3}\footnotesize
\textit{Note.} Academic performance includes 5-year normalized citation scores and the number of first- and last-authored publications. Collaboration networks are represented by the average impact of coauthors (measured by their 5-year citation scores prior to copublication), number of coauthors, and proportion of female coauthors. Institutional background is captured by the impact of institutional affiliation (proxied by the total 5-year citation scores of affiliated publications), location of the original affiliation, and the number of unique affiliations held. Except for the origin region of affiliations (shown as aggregate proportions), all metrics are calculated annually for each researcher by academic age (x-axis) and aggregated to show average trends of each measurement (y-axis) by gender. Note that ``North America'' here refers to the United States and Canada, excluding Mexico. Shaded areas represent 95\% confidence intervals. No.\ = number.
\end{minipage}
\end{figure}

To examine potential drivers of gendered differences in academic progression, Figure~\ref{fig:var_stat} presents descriptive trends across three domains: academic performance, collaboration networks, and institutional affiliation. These indicators are tracked by academic age and disaggregated by gender to illustrate how women's and men's career profiles evolve over time. All indicators are derived from publication records.

Men had an early-career advantage based on conventional metrics of academic performance. The 5-year normalized citation score combines productivity and impact by first normalizing the citations received within 5 years of each publication by discipline and publishing year and then aggregating these normalized values across all publications from the same year. The aggregated citations, labeled as citation scores, increased over time for both women and men; however, men’s scores rose more steeply, leading to a performance gap that peaked around the academic age of 7. This gap later narrowed as citation growth rates converged. A similar pattern appeared in first-authored publications: While output rose for both genders from a low baseline, the increase for men was sharper, resulting in a peak gap of about 0.1 publications at academic age 7 (roughly one extra article per 10 male researchers) before the first-authored output for both genders declined. Men consistently maintained an advantage in last-authored publications (a proxy for seniority) throughout their careers.

Early in their careers, women tended to collaborate with slightly more influential researchers than did men, though the impact of collaborators declined for both genders over time, likely reflecting a natural shift toward mentoring junior researchers. As careers progressed, women and men both expanded the number of their coauthors, but gender differences emerged in network composition. Women began with gender-balanced collaborations that became increasingly female dominated, whereas men started with male-dominated networks that gradually became more balanced with academic age. 

While both women and men tended to affiliate with higher impact institutions during early and senior career stages (and relatively lower impact ones in midcareer), women were affiliated with more prestigious institutions than men across most career stages. Geographically, over 80\% of psychology entrants began their careers in North America (excluding Mexico)  and Europe, with women especially likely to start at North American institutes. Men had higher mobility, which is a well-recognized strategy for career advancement: They accumulated slightly more unique affiliations than women on average (4.0 vs. 3.6). However, when disaggregated by academic age, women and men exhibited a similar pattern of steadily increasing affiliations before academic age 15, suggesting that men’s longer careers largely account for their higher total number of affiliations.
Gender differences in academic profiles, disaggregated by the three cohort groups (2000–2004, 2005–2009, and 2010–2014), are shown in Supplemental Figure~S2. 

In sum, we found a clear divergence in career profiles: Men performed better on traditional metrics like citations and first authorships in their early careers, while women had more prestigious collaborators and institutions.

\subsection*{Academic Performance Is the Strongest Predictor of Attrition}\label{result3}

To identify which factors were most strongly related to academic attrition and how their relevance changed across the career, we incorporated the core factors of academic performance (citation scores, number of first-authored articles, and number of last-authored articles), collaboration (coauthor impact, number of coauthors, and proportion of women coauthors), and institutional affiliation (affiliation scores, origin region of affiliation, and number of unique affiliations) into a time-varying effect model (see more details in the Supplemental Materials). This approach allows the association between each factor and attrition risk to change over the course of an academic career rather than assuming constant effects. We included psychology subfields to account for disciplinary variation. Initial subfield and geographic region of first affiliation were treated as time-invariant predictors, whereas performance, collaboration, and institutional variables were modeled as time-varying predictors. Gender was also specified as a time-varying effect, enabling us to assess whether its association with attrition differed across career stages. 

Figure~\ref{fig:importance} shows the results (detailed estimate results are provided in Supplemental Tables~S2 and S3). For each cohort group, the left-hand panel shows the overall importance of each factor in predicting attrition across all career stages. Higher values indicate a stronger contribution to the model's prediction of whether a researcher exits academia. The corresponding heatmap on the right illustrates how the influence of each time-varying factor changes across academic age, highlighting the timing of these effects. While the overall importance captures the average contribution across the full career, the heatmaps reveal that some effects are concentrated at specific academic stages and may offset or diminish over time, resulting in a more modest aggregate impact. Blue indicates higher retention, while red indicates higher attrition.

\begin{figure}[!htbp]
\centering
\caption{Relative Importance of Predictors for Academic Attrition and Their Temporal Dynamics}
\label{fig:importance}
\includegraphics[width=\textwidth]{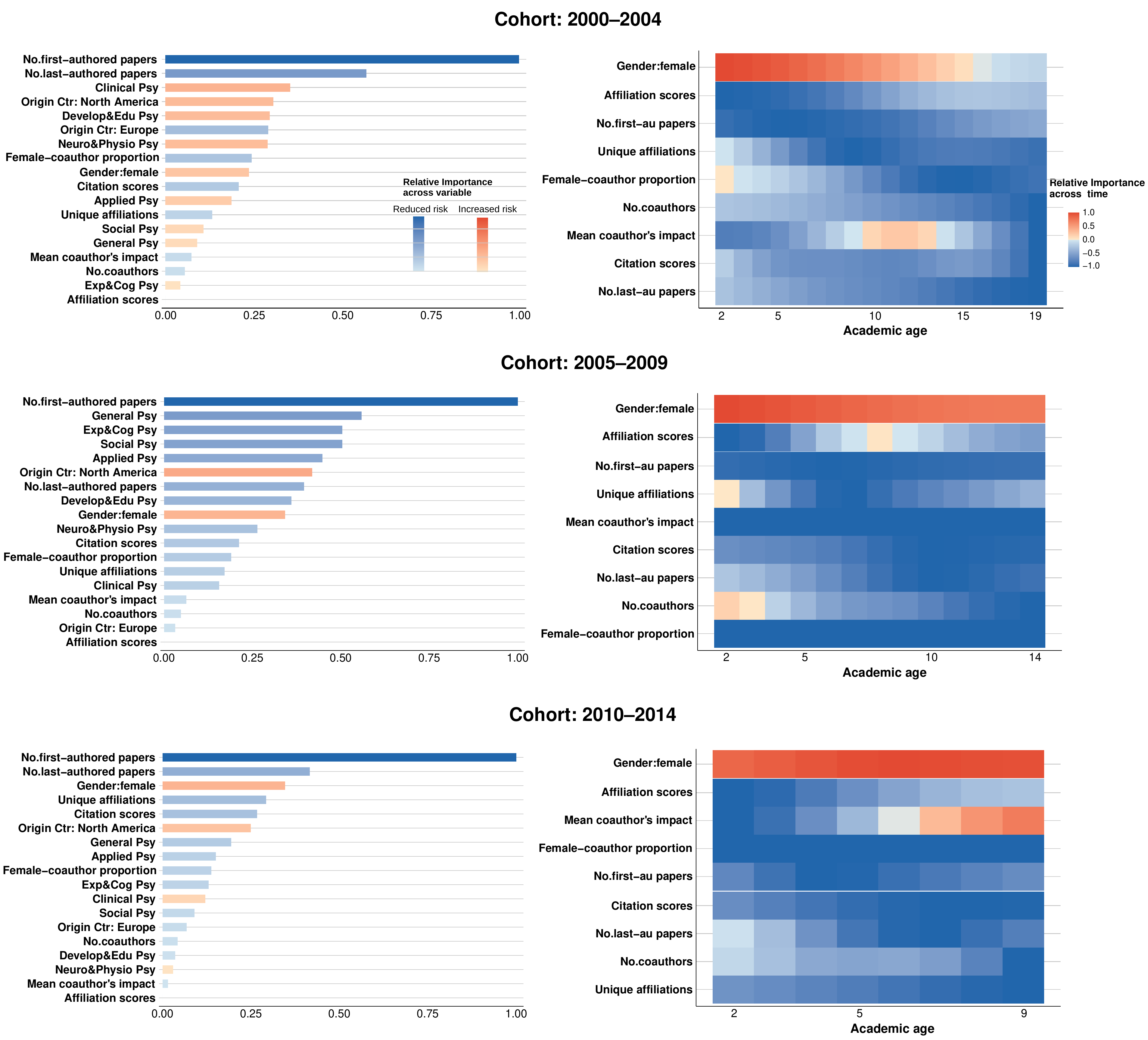}

\vspace{6pt}
\begin{minipage}{\textwidth}
\renewcommand{\baselinestretch}{1.3}\footnotesize
\textit{Note.} Blue (red) indicates a decreasing (increasing) attrition risk over time. Model coefficients were scaled to a 0--1 range, where higher values indicate greater influence on attrition risk. Left: Bar plots show the relative importance of time-varying and time-variant predictors from the cohort groups of 2000--2004, 2005--2009, and 2010--2014. Predictors are ordered by magnitude. Right: Heatmaps show how the relative importance of time-varying predictors evolved across academic ages. Note that the earliest cohort has the longest career trajectory due to data availability. Predictors are ordered by the academic age at which they are most influential: Those appearing higher (lower) tended to shape early-career (late-career) outcomes. No.\ = number; Develop \& Edu Psy = developmental and educational psychology; Neuro \& Physio Psy = neuroscience and physiological psychology; Clinical Psy = clinical psychology; General Psy = general psychology; Exp \& Cog Psy = experimental and cognitive psychology; Social Psy = social psychology; Applied Psy = applied psychology.
\end{minipage}
\end{figure}

Across cohort groups, academic performance was the strongest predictor of career retention. The number of first-authored publications consistently ranked as the most important factor, with its influence peaking during the early-career phase (around academic ages 4–5) indicated in the left panel. The number of last-authored publications, a proxy for seniority, was the second most important factor, becoming increasingly influential in later-career stages. While higher citation scores were linked to greater retention, their predictive power was strongest for the most recent cohort group (2010–2014), suggesting a growing emphasis on citation impact over time.

Collaboration network characteristics were influential, but their effects varied across career stages. For example, in the earliest cohort, a higher proportion of female coauthors was associated with increased attrition risk early in the career but with reduced risk at later stages. Among institutional factors, institutional prestige showed a modest positive association with retention early in the career, but its overall influence remained limited. Notably, beginning a career in North America was modestly associated with a higher risk of attrition, particularly in earlier cohorts.

We confirmed the robustness of these findings with a model that used lagged variables to account for potential confounding between a predictor and the publication-based survival metric. The results were consistent with our main analysis (see Supplemental Figure~S3).

\subsection*{Elevated Female Attrition for Equal Performance, Collaboration, and Affiliation}

\begin{figure}[htbp!]
\centering
\caption{Predicted Attrition by Gender Across Academic Age and Cohort Group}
\includegraphics[width=1.02\textwidth]{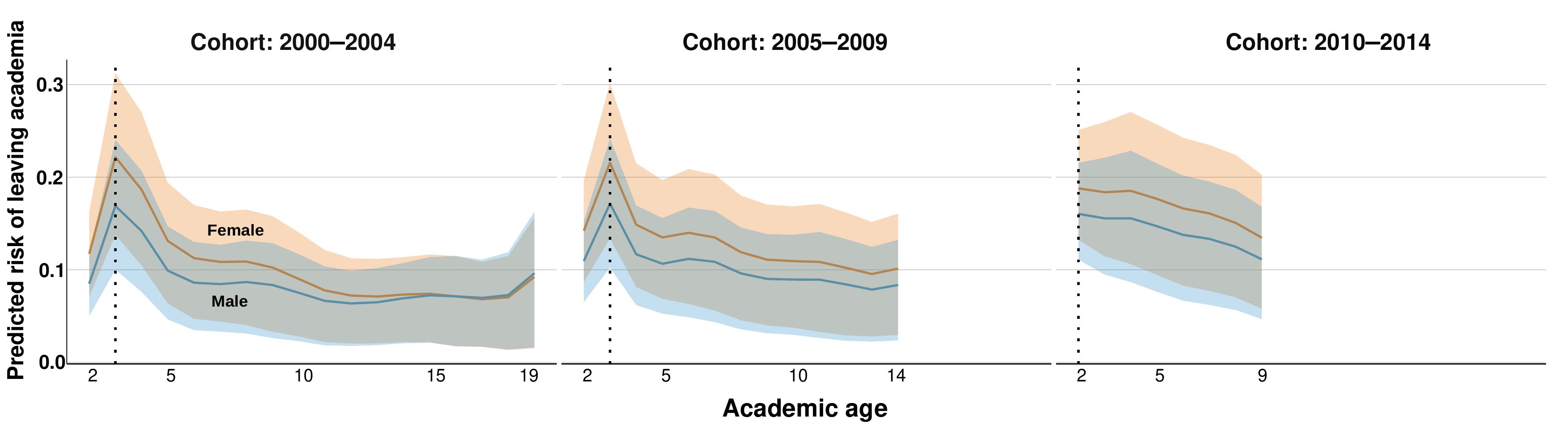}
\label{fig:prediction}
\vspace{6pt}
\begin{minipage}{\textwidth}
\renewcommand{\baselinestretch}{1.3}\footnotesize
\textit{Note.} The predictions average marginal contrasts derived from the time-varying effects model, holding all other covariates constant while varying only gender. Shaded bands indicate the interquartile range (25th--75th percentile) of the predicted values. Results are shown separately for researchers entering academia in the 2000–2004, 2005–2009, and 2010–2014 cohort groups.
\end{minipage}
\end{figure}

The gender differences in academic profile, along with their impact on attrition, raise a key question: Can these factors fully explain gender disparities in academic progression? In other words, if women and men had comparable academic profiles, would they face the same attrition risks across their careers? To test this, we predicted the risk of attrition, that is, the cessation of publishing, by academic age while holding performance, collaboration, and institutional factors constant, based on the result from time-varying effect model described above (see Figure~\ref{fig:prediction}). Shaded bands represent confidence intervals, capturing the uncertainty around predictions for each gender. While the intervals are relatively wide, particularly at later-career stages due to smaller sample sizes, overlap between bands does not necessarily imply a lack of statistical significance. Indeed, bootstrap results from the model-based contrasts (see Supplemental Figure~S4) show that gender gaps across cohort groups remain significantly persistent. This indicates that women face a higher risk of attrition than their male peers with comparable academic profiles, with the gap most pronounced during the high-risk early-career years.

The pattern of the disparity between female and male researchers shifted over time. In the earliest cohort group (2000–2004), the gender gap in the predicted attrition risk peaked by academic age 3, at 5\% (i.e., the predicted attrition risk was over 20\% for women vs. 15\% for men). The gap then narrowed with the decline in attrition for both genders and eventually closed by academic age 15. The 2005–2009 cohort group showed a similar pattern but with higher overall attrition and a persistent 2\% gender gap. For the most recent cohort group (2010–2014), the gender gap emerged earlier (academic age 2) and remained stable at around 3\% through academic age 9, suggesting that the disparity did not diminish over time. Additional results by region of affiliation are presented in Supplemental Figure~S5.

Ultimately, these findings show that a substantial portion of the gender gap in academic careers cannot be attributed to bibliometrically measurable differences in performance or prestige. This persistent disparity, especially in recent cohorts, points toward the influence of broader structural or social factors that systematically disadvantage women in academia. Even when women and men have statistically comparable records of performance, collaboration, and prestige, women still face a significantly higher attrition risk in publishing.

\section*{Discussion}\label{sec12}

By tracing the careers of over 78,000 researchers worldwide, this study provides a comprehensive, longitudinal analysis of gendered attrition and retention in academic psychology and evaluates the role of key career metrics. Despite women outnumbering men at entry across all psychology subfields, they remain underrepresented in senior academic ranks \parencite{Gruber_psychology_2020,Casad2022_SocialScience}. By tracing the career progression of researchers who entered the field between 2000 and 2014 and observing their outcomes up to 2024, we demonstrate that women consistently face higher attrition risks in publishing, especially during the early-career phase. We confirm previous evidence of a gender gap in early-career first-authored publications \parencite{Odic_pubgap_2020} and extend this literature by demonstrating its importance for long-term career retention. Even after accounting for performance, collaboration, and institutional context, women still faced higher attrition risks in the early stages of their careers, while the gender gap closed in late-career stages.

These findings challenge the narratives that underrepresentation is primarily driven by recruitment deficits and that gender gaps in early-career advancement in psychology appear to have closed. Instead, our findings highlight the cumulative impact of postentry mechanisms of attrition \parencite{Blackburn_stem_2017, EC_shefigure_2021}. 

Across the span of academic careers in psychology, our analyses indicate that being a woman is consistently associated with a higher risk of attrition. These findings diverge from studies reporting small or narrowing gender gaps in tenure attainment or promotion rates \parencite{Gruber_psychology_2020, Janet_survival_social_2015, webber2018there}. However, those studies typically rely on survey-based or institution-level data from a limited set of countries, most commonly the United States, and focus on specific promotion transitions, whereas our bibliometric measure captures sustained research engagement across entire career trajectories and a broader international context. The persistence of gender attrition gaps points to the presence of structural or unmeasured barriers to academic progression, particularly during the early stages of an academic career, when overall attrition risk is at its highest. The eventual narrowing of gender gaps at later stages may reflect selective retention, whereby only a subset of women with exceptional support, resilience, or opportunity remain actively publishing. It may also reflect a waning burden of childcare and family responsibilities over time, as these responsibilities disproportionately disadvantage women \parencite{goulden2011keeping,Mason_2013_baby}. These dynamics likely extend to other fields in the social and behavioral sciences, which may mirror a similar mismatch between high female entry and low senior representation \parencite{Ceci_causes_2011, Habicht2026_leaky_pipeline}. Our findings broaden the conversation on gender inequality in academia, demonstrating that postentry attrition, rather than recruitment or admission alone, is a pervasive and enduring challenge that must be addressed.

One pathway through which this attrition operates is differential academic performance. Women are underrepresented in key metrics of academic performance, including publication, collaboration, and institutional affiliation---a pattern long recognized as contributing to higher attrition risk \parencite{Ceci_causes_2011, van2012gender, Larivire_gender_2013, Huang_historical_2020}. Our findings confirm earlier evidence of gender differences in productivity in psychology \parencite{Odic_pubgap_2020}. However, we additionally show how career metrics contribute to academic retention and that gender gaps are not static, narrowing in later-career stages and varying across cohorts. Disparities in first and last authorships may reflect unequal access to resources and opportunities, as well as unequal family responsibilities, particularly during early-career stages. Consistent with prior work on coauthorship patterns \parencite{Odic_pubgap_2020} and institutional placement \parencite{Way2016_gender}, women perform comparably, and in some cases more strongly, in measures of collaboration and affiliations. However, compared with publication performance, collaboration is associated with only a small reduction in attrition risk, and institutional prestige exerts little influence on retention. Importantly, these effects are not constant across the career: for example, the impact of coauthor characteristics becomes more positive during midcareer stages. Taken together, our findings suggest that women’s lower performance in certain aspects of early-career stages may rather stem from unequal structural conditions than lower institutional competitiveness. 

A key mechanism that may contribute to these early-career gender differences, alongside disparities in hiring, funding, and promotion, is the timing of childbearing relative to academic career progression \parencite{Mason_2013_baby,Paksi_2016_timing,Sugimoto_2023_WomenBook, Nakagawa_tenure_2024}. Entry into independent academic positions often coincides with ages at which many women face increasing biological and social pressures around family formation, while early-career stages are also characterized by high productivity demands and evaluation pressures \parencite{Mason_2013_baby, wolfinger2008problems, Wolfinger_2009_stay}. These overlapping constraints may make it more difficult to sustain continuous academic productivity during this critical period, particularly for those experiencing pregnancy, childbirth, or early childcare responsibilities. Our data do not include direct measures of parenthood or fertility timing, and therefore, we cannot directly test this mechanism. Nonetheless, we observed that the attrition gap largely dissolves at later-career stages. Our findings should thus not be interpreted as reflecting differences in ability but rather as consistent with structural and life-course constraints that likely differentially shape career trajectories.

Our findings suggest several practical steps that institutions and funding bodies could take to better support academic retention of women in psychology. Our analysis identifies the early-career stage as a particularly vulnerable period for female researchers,  suggesting that targeted support during this phase could play a crucial role in reducing gender disparities. Interventions could include expanding access to mentorship, increasing the visibility of role models, ensuring equitable recognition in authorship and evaluation processes, and offering greater flexibility to accommodate competing demands \parencite{Gruber_psychology_2020, Fernandes2023_program,Mason_2013_baby}, particularly during major transitions such as moving from a PhD to a postdoctoral role or from a postdoctoral role to a tenure-track position. For instance, the National Institutes of Health Early Career Reviewer Program provides structured mentorship opportunities, and European funding bodies such as the European Research Council allow extensions to accommodate parental leave. Building on and broadening such measures will be critical for supporting more equitable career progression.

Targeted funding schemes or fellowships for women, as well as flexibility in age or career-stage criteria for such positions, may help address disadvantages faced by those with caregiving responsibilities or who may anticipate these responsibilities in the near future. Similar measures may apply to other social science disciplines that face comparable attrition patterns as psychology; however, the generalizability of our findings to other disciplines remains to be tested.
 
However, the reasons for gender disparities in research careers are highly complex, with many additional factors likely influencing attrition beyond those captured by our bibliometric measures. Other crucial contextual influences include mentorship quality, institutional climate, and individual preferences. Mixed methods approaches are therefore needed to capture the full range of mechanisms underlying attrition gaps and to inform the development of effective interventions. 

\subsection*{Constraints on Generality}

Given the methodological approach and the large number of psychology researchers in our database, we have no reason to believe that the sample is unrepresentative of the population of researchers in psychology who pursue an academic career. Methodologically, our study demonstrates the value of large-scale bibliometric analysis for studying academic career trajectories. Unlike survey-based approaches, this approach enables robust, longitudinal tracking of entire research populations, capturing the dynamic influence of multiple factors on retention \parencite{Fortunato2018_scisci, Huang_historical_2020}. Moreover, the ability to compare gendered career outcomes across global contexts adds an international lens to a literature that is often limited to U.S. institutions \parencite{webber2018there, Janet_survival_social_2015}. The use of an international name-based gender inference enables the study of gender disparities at this scale and extends similar efforts using U.S. or U.K. census data \parencite{Odic_pubgap_2020}, even though some measurement bias remains unavoidable. Overall, the scale, depth, and coverage of bibliometric data offer powerful insights into the dynamics of systemic inequality in science.

At the same time, our bibliometric approach carries specific interpretive limits. Our sample captures researchers who have entered the publication system and therefore does not include individuals who pursued academic pathways (e.g., completed doctoral training) but exited before producing their first indexed publication. Similarly, our operationalization of attrition as cessation of publishing activity captures disengagement from active research output but may not fully distinguish researchers who leave academic positions entirely from those who remain in academic-adjacent roles (e.g., teaching-focused positions, administration, or applied work) without continuing to publish. Future work combining bibliometric and employment data could further disentangle these pathways.

Similarly, bibliometric data do not include personal information on parenthood, fertility timing, or family leave, which prevents us from directly incorporating childbearing into our statistical models. Given the role that the timing of family formation may play in shaping early-career trajectories, particularly for women, linking bibliometric career data with parenthood records or survey-based measures of family formation is therefore a priority for future research. 

Finally, author disambiguation and gender inference, while advanced, are imperfect. Cultural practices surrounding marital name changes and other forms of name variation over time \parencite{Santamara2018_namegender,Zhao2023_migration} complicate author disambiguation. Although we implemented large language model-assisted strategies to detect name changes, these complexities cannot be fully resolved. Additionally, name-based gender prediction cannot capture the identities of nonbinary or transgender researchers, and it inevitably misclassifies the gender of a small subset of researchers. Future research must continue to develop more inclusive and nuanced data practices to better represent the full spectrum of identities in academia.

\section*{Conclusion}
Taken together, by tracing the careers of over 78,000 researchers in the field of psychology worldwide, we show that women are disproportionately more likely to exit active research careers, particularly during the early-career stage, and that this pattern has not improved over time. These disparities are not fully explained by differences in research performance, collaboration, or institutional prestige, highlighting the complexities underlying higher attrition rates for women. Our findings suggest that even when women enter psychology at higher rates than men, ensuring their long-term retention and advancement is a critical challenge to facilitate gender parity across career trajectories.

While studying inequality in science and society is increasingly politicized, it is vital to ground discussions in robust empirical evidence. Large-scale, career-long analyses can help identify when gender gaps emerge and where institutional interventions are most effective. The present findings can, therefore, inform constructive policy and organizational practices that foster equitable and sustainable academic careers for all.

\newpage
\printbibliography

\end{document}